\documentclass[eqsecnum,preprint,prd,aps,nofootinbib]{revtex4}
\usepackage{amsmath}
\usepackage{graphics}
\newcommand{\be}{\begin{equation}}
\newcommand{\ee}{\end{equation}}
\newcommand{\ba}{\begin{eqnarray}}
\newcommand{\ea}{\end{eqnarray}}
\newcommand{\bc}{\begin{center}}
\newcommand{\ec}{\end{center}}

\begin{document}
\begin{center}
\bibliographystyle{article}

{\Large \textsc{Noether symmetry approach in matter--dominated
cosmology with variable $G$ and $\Lambda$}}

\end{center}
\vspace{0.4cm}

%\vspace*{0.6cm}

\date{\today}

\author{Alfio Bonanno,$^{1,2}$ \thanks{%
Electronic address: abo@ct.astro.it} Giampiero Esposito$^{3,4}$ \thanks{%
Electronic address: giampiero.esposito@na.infn.it} Claudio Rubano$^{4,3}$
\thanks{%
Electronic address: claudio.rubano@na.infn.it} and Paolo Scudellaro$^{4,3}$
\thanks{%
Electronic address: scud@na.infn.it}}
\affiliation{${\ }^{1}$Osservatorio Astrofisico, Via S. Sofia 78, 95123 Catania, Italy\\
${\ }^{2}$Istituto Nazionale di Fisica Nucleare, Sezione di Catania,\\
Corso Italia 57, 95129 Catania, Italy\\
${\ }^{3}$Istituto Nazionale di Fisica Nucleare, Sezione di Napoli,\\
Complesso Universitario di Monte S. Angelo, Via Cintia, Edificio N', 80126
Napoli, Italy\\
${\ }^{4}$Dipartimento di Scienze Fisiche, Complesso Universitario di Monte
S. Angelo,\\
Via Cintia, Edificio N', 80126 Napoli, Italy}

\begin{abstract}
In the framework of renormalization-group improved cosmologies, we use
the Noether symmetry approach to get exact and general integration of the
matter--dominated cosmological equations. This is performed by using
an expression of $\Lambda = \Lambda (G)$ determined by the method
itself. We also work out a comparison between such a model and the
concordance $\Lambda$CDM model as to the magnitude--redshift
relationship, hence showing that no appreciable differences occur.
\end{abstract}
\maketitle
\bigskip
\vspace{2cm}

\section{Introduction}

The study of cosmological dynamics has been recently performed by
analyzing strong ``renormalization group (RG) induced'' quantum effects.
These are supposed to drive the (dimensionless) cosmological
``constant'' $\lambda(k)$ and Newton ``constant'' $g(k)$ from an
ultraviolet attractive fixed point. This approach acts within the
framework of the effective average action
\cite{Berg02,Wett01,Reut94}, finding a fixed point in the
Einstein--Hilbert truncation of theory space
\cite{Reut98,Laus02a,Soum99} and in the higher--derivative
generalization \cite{Laus02b}. The existence of such a
non-Gaussian ultraviolet fixed point in the exact theory implies its
nonperturbative renormalizability
\cite{Laus02a,Reut02b,Laus02b,Bona05,Nied03,Nied02,Forg02}. This
RG--improved framework describes gravity at a typical distance
scale $\ell\equiv k^{-1}$, introducing an effective average action
$\Gamma_k[g_{\mu\nu}]$ for Euclidean quantum gravity
\cite{Reut98}, and deriving an exact functional RG equation for
the $k$--dependence of $\Gamma_k$. This context is usually
referred to as \emph{quantum Einstein gravity}, and makes it
possible to find an explicit $k$--dependence of the running Newton
term $G(k)$ and the running cosmological term $\Lambda(k)$. This
is interesting for an understanding of the Planck era immediately
after the big bang as well as the structure of black hole
singularity \cite{Bona02b,Bona99,Bona00}.

In order to obtain the RG--improved Einstein equations for a
homogeneous and isotropic universe, one can identify $k$ with the
inverse of cosmological time, $k \propto 1/t$
\cite{Bona02b,Bona02a}, and a dynamical evolution for $G$ and
$\Lambda(k)$ induced by their RG running can thus be derived. An
Arnowitt--Deser--Misner (ADM) formulation has also been presented
\cite{Bona04}; it builds a modified action functional which
reduces to the Einstein--Hilbert action when $G$ is constant. A
power--law growth of the scale factor can then be obtained for
pure gravity and for a massless $\varphi^4$ theory in a
homogeneous and isotropic space--time, in agreement with what is
already known on fixed--point cosmology. On the other hand, still
within the framework of the ADM formulation, in \cite{Bona06} we
have proposed solutions for the pure gravity case derived by means
of the so-called \emph{Noether Symmetry Approach}. This is a method
which implements a variable transformation that usually makes it possible
to find exact and general solutions of the cosmological equations
\cite{deritis90,cap96}. The solutions found in Ref. \cite{Bona06}
predict that the Universe is in an accelerated stage, hence mimicking
inflation without introducing a scalar field in the cosmic
content.

There are also some indications \cite{Tsam93} that quantum
Einstein gravity, because of its inherent infrared divergences, is
subject to strong renormalization effects even at very large
distances. In cosmology, such effects might be relevant for the
Universe at late times, since they lead to a dynamical relaxation
of $\Lambda$ and, probably, may solve the cosmological constant
problem \cite{Tsam93}. As a matter of fact, the late accelerated
expansion of the Universe can be viewed as a renormalization group
evolution near a non--Gaussian infrared fixed point
\cite{Bona02a}, where $G$ and $\Lambda$ become running quantities
at some late time. The sharp transition between standard FLRW
cosmology and accelerated RG driven expansion is supposed to occur
at some time (the fixed point being reached exactly at the
transition), but is a strong simplifying assumption; however,
some agreement can be found between this kind of model and
SnIa observations \cite{Bent04}. In what follows, we would like
to present some exact solutions of the improved Hamiltonian cosmology
described in \cite{Bona04}, {\it without} any assumption on 
fixed-points structure, discussing the possible physical viability
of the resulting model. Interestingly, it is possible to think that the 
kind of `post-inflationary' accelerated stage we discover in our
considerations could explain what is observed by means of SnIa
data \cite{perl1,perl2,riess,garn}. We stress again that we do not
assume a priori what functional relation exists between $\Lambda$
and $G$, but we merely postulate its occurrence. Some solutions compatible
with having $\Lambda G= {\rm constant}$ were found in the ultraviolet 
regime in \cite{Bona04}, whereas we here find a different 
renormalization-group trajectory. 

Here, we investigate the matter--dominated universe, limiting our
analysis to cosmology in the dust case. Our aim is to derive the
main parameters that make it possible to compare our model with the
concordance $\Lambda$CDM model as to the magnitude--redshift
relationship, hence showing that no appreciable differences indeed
result when one wants to examine Type-Ia supernovae data. 
In section 2 we describe the Lagrangian formulation for
the Lagrangian adopted to derive the RG--improved Einstein
cosmological equations with the ordinary matter energy--momentum
tensor and then find the Noether symmetry. Section 3 studies how
this gives rise to exact and general solutions, while in section 4
we discuss the situation with ${\cal K}=0$, comparing our model
and the $\Lambda$CDM model. In section 5 we draw conclusions.

\section{Noether symmetry}

First of all, let us consider the approach outlined in Ref.
\cite{Bona04} and there applied to models of gravity with variable
$G$ and $\Lambda$. From the modified action functional, an
Einstein--Hilbert action can in fact be obtained when $G$ and
$\Lambda$ are such. As said, in a homogeneous and isotropic
universe, this leads to power--law behaviours of the cosmic scale
factor $a = a(t)$ for both pure gravity and a massless $\varphi^4$
theory, in agreement with results from fixed-point cosmology, once
one adopts the constraint $G\Lambda = const$. On the other hand,
it is known that an independent dynamical $G$ is equivalent to
metric-scalar gravity already at classical level
\cite{Cap97,Cap98}. Taking both $G$ and $\Lambda$ as varying
independently with position and time can drive to pathological
situations, and it is possible to show that we must therefore
assume $\Lambda = \Lambda(G)$ \cite{Bona04}.

Here, we want to investigate the matter--dominated case in
homogeneous and isotropic cosmology (with a signature $-,+,+,+$
for the metric, lapse function $N=1$ and shift vector $N^{i}=0$). 
As in Ref. \cite{Bona04} (but see also Ref.
\cite{Bona06}), let us start from the Lagrangian
\be \label{2} L =
\frac{1}{8 \pi G} \left( -3a\dot{a}^2 + 3{\cal K}a - a^3 \Lambda +
\frac{1}{2}\mu a^3 \frac{\dot{G}^2}{G^2} \right) -
Da^{-3(\gamma-1)}\,,
\ee
where $G=G(t)$, $\Lambda=\Lambda(G(t))$,
${\cal K} = -1,0,1$ (for open, spatially flat and closed
universes, respectively), while dots indicate time derivatives and
$\mu$ is a nonzero interaction parameter introduced in Ref.
\cite{Bona04} and also used in Ref. \cite{Bona06}. We have now
inserted also the matter contribution by means of $L_m \equiv -
Da^{-3(\gamma-1)}$, with $1 \leq \gamma \leq 2$ (being, in the
most relevant cases, $\gamma = 1$ for dust and $\gamma = 4/3$ for
radiation) and D a suitable integration constant connected to the
matter content. From Eq. (\ref{2}) we find the second-order
Euler--Lagrange equations for $a$ and $G$
\be \label{3}
\frac{\ddot{a}}{a} + \frac{\dot{a}^2}{2a^2} + \frac{{\cal
K}}{2a^2} - \frac{\Lambda}{2} - \frac{\dot{a}\dot{G}}{a G} +
\frac{\mu \dot{G}^2}{4G^2} + 4\pi G(\gamma-1)Da^{-3\gamma}=
0\,,\ee \be \label{4} \mu \ddot{G} - \frac{3}{2}\mu
\frac{{\dot{G}}^2}{G} + 3\mu \frac{\dot{a}}{a}\dot{G} +
\frac{G}{2} \left( -6\frac{\dot{a}^2}{a^2} + \frac{6{\cal
K}}{a^2}-2\Lambda + 2G\frac{d\Lambda}{d G} \right) = 0\,.
\ee
We have also to consider the Hamiltonian constraint \cite{Bona04}
\be
\label{5} \frac{\dot{a}^2}{a^2} + \frac{{\cal K}}{a^2} -
\frac{\Lambda}{3} - \frac{\mu}{6}\frac{\dot{G}^2}{G^2} -\frac{8\pi
G}{3}Da^{-3\gamma}= 0\,, \ee which can indeed be seen as
equivalent to the following constraint on the \emph{energy}
function associated with $L$ \cite{deritis90,cap96,Bona06}: \be
\label{5bis} E_L \equiv \frac{\partial L}{\partial \dot{a}}\dot{a}
+ \frac{\partial L}{\partial \dot{G}}\dot{G} - L = 0\,.
\ee

In what follows we want to discuss the dust case, so choosing
$\gamma = 1$; this involves a zero pressure $p_m$ and an energy
density $\rho_m = Da^{-3}$. The Lagrangian $L$ takes therefore
the simplified form
\be \label{5bis1} L = \frac{1}{8 \pi G} \left(
-3a\dot{a}^2 + 3{\cal K}a - a^3 \Lambda + \frac{1}{2}\mu a^3
\frac{\dot{G}^2}{G^2} \right) - D\,,
\ee
so that the matter term
is now just a constant and has no effect on the equations of
motion with respect to the pure gravity case, but it is
nonetheless important, since it always occurs in the constraint
equation (\ref{5bis}). We have already solved the system of
equations of motion in pure gravity \cite{Bona06}, using the
Noether Symmetry Approach \cite{deritis90,cap96}. Therefore, let
us again consider the Lagrangian $L$ as a \emph{point-like}
Lagrangian, function of the variables $a$ and $G$, 
and their first derivatives \cite{deritis90,cap96,Bona06}. In Ref.
\cite{Bona06}, we have shown that, with a consistent choice of the
function $\Lambda = \Lambda(G)$, a Noether symmetry indeed exists
for the pure gravity Lagrangian. A key point in our considerations
is now given by the observation that, in the matter--dominated
case, the procedure to deduce the symmetry is exactly the same as
in the pure gravity situation, the only substantial difference
being in Eq. (\ref{5bis}) (which is in fact equivalent, now, to
Eq. (\ref{5})), therefore acting as a slightly different (but very
important) constraint on the integration constants involved by the
solution method. This means that we find exactly the \emph{same}
Noether symmetry as in \cite{Bona06}, so that we can here use the
same transformations introduced there, and write the expressions
for $a=a(t)$ and $G=G(t)$ obtained there, but taking care to
consider, now, the updated energy constraint.

More in detail, let us consider the vector field
\be \label{6} X
\equiv \alpha (a,G)\frac{\partial{}}{\partial a} +
\dot{\alpha}\frac{\partial{}}{\partial \dot{a}} + \beta
(a,G)\frac{\partial{}}{\partial G} +
\dot{\beta}\frac{\partial{}}{\partial \dot{G}}\,,
\ee
with $\alpha
= \alpha (a,G)$ and $\beta = \beta (a,G)$ generic $C^{1}$
functions, and $\dot{\alpha} \equiv d\alpha/dt = (\partial
\alpha/\partial a)\dot{a} + (\partial \alpha/\partial G)\dot{G}$,
$\dot{\beta} \equiv d\beta/dt = (\partial \beta/\partial a)\dot{a}
+ (\partial \beta/\partial G)\dot{G}$. As in Ref. \cite{Bona06},
the condition
\be \label{7} {\cal L}_X L = 0
\ee
(${\cal L}_X L$
being the Lie derivative of $L$ along $X$) corresponds to a set of
equations for $\alpha = \alpha (a,G)$, $\beta = \beta (a,G)$ and
$\Lambda = \Lambda (G)$, which can be solved by \cite{Bona06}
\be
\label{7bis} \alpha (a,G) = a^{\frac{J}{3-2J}}G^{J-1}\,,\,\,\,\,\,
\beta (a,G) = \frac{3}{3-2J}a^{\frac{3(J-1)}{3-2J}}G^J\,.
\ee
The $J$ parameter is an arbitrary constant $\neq 1, 3/2$ and such that
\be \label{13bis}
\mu = \frac{2}{3}(3-2J)^2 \neq 0, \frac{2}{3}\,.
\ee
(In what follows we prefer to use $J$ instead of $\mu$ so as to
simplify the resulting expressions.) For consistency, we still
get both
\be \label{15}
J{\cal K} = 0
\ee
and
\be \label{16}
2(1-J)\Lambda + G\frac{d\Lambda}{d G} = 0\,,
\ee
which splits our considerations into two separate branches.

Thus, neglecting unnecessary integration constants, we have 
\ba
X_J & \equiv & X|_{\forall J\neq 0} =
a^{\frac{J}{3-2J}}G^{J-1}\frac{\partial{}}{\partial a} +
aG^{J-2}\left[ \frac{J}{3-2J}a^{\frac{5J-6}{3-2J}}G\dot{a} +
(J-1)a^{\frac{3(J-1)}{3-2J}}\dot{G} \right]
\frac{\partial{}}{\partial \dot{a}} \nonumber \\  &  & +
\frac{3}{3-2J}a^{\frac{3(J-1)}{3-2J}}G^J\frac{\partial{}}{\partial
G} + \frac{3}{3-2J}a^{\frac{5J-6}{3-2J}}G^{J-1}\left[
\frac{3(J-1)}{3-2J}G\dot{a} + Ja\dot{G} \right]
\frac{\partial{}}{\partial \dot{G}}, \ea \be \label{18}
\Lambda=\Lambda(G)=W G^{2(J-1)}, 
\ee 
for $J\neq 0, 1, 3/2$ ($\Rightarrow$ any
$\mu \neq 0, 2/3$) and ${\cal K}=0$. It turns out that $X_J
\rightarrow X_0$ (as well as for the expression of $\Lambda =
\Lambda (G)$) for $J \rightarrow 0$, but the situation with $J=0$
($\Rightarrow \mu = 6$) has to be nonetheless treated separately,
since it is not a simple subcase except when ${\cal K}=0$.

\section{Solutions from new coordinates and Lagrangian}

There exists a change of coordinates $\{a,G\} \rightarrow\{u,v\}$,
such that one of them (say $u$, for example) is cyclic for the
Lagrangian $L$, and the transformed Lagrangian produces exactly
and generally solvable equations. Solving the system of equations
$i_{X} d u = 1$ and $i_X d v = 0$ (where $i_X d u$ and $i_X d v$
are the contractions between the vector field $X$ and the
differential forms $d u$ and $d v$, respectively
\cite{deritis90,cap96}), we get
\be \label{18bis}
u = u(a,G) = n\,
a^{\frac{1}{2n}} G^m\,,\,\,\,\,\, v = v(a,G) = \ln{\left( a G^{-2n
m} \right)}\,.
\ee
Here, we have defined
\be \label{3.5}
n \equiv
n(J) \equiv \frac{3-2J}{6(1-J)}\,,\,\,\,\,\,  m \equiv m(J) \equiv
1-J = \frac{1}{2(3n - 1)}\,,
\ee
which are well defined and
non-vanishing, being $J \neq 1, 3/2$. Eq. (\ref{18}) thus becomes
\be \label{18bis1}
\Lambda = \Lambda (G) = W
G^{\frac{1}{1-3n}} \,.
\ee
In the case $J=0$ (i.e., $n = 1/2$ and
$m=1$) and any ${\cal K}$, we then get
\ba u = u(a,G) & = &
\frac{1}{2}a G\,, \\ v = v(a,G) & = & \ln{\left( a G^{-1}
\right)}\,.
\ea

The inversion of Eq. (\ref{18bis}) gives
\ba a = a(u,v) & = &
n^{-n}
\exp{\left( \frac{v}{2} \right)} u^n\,, \\
G = G(u,v) & = & {\left[ \frac{1}{n} \exp{\left( - \frac{v}{2n}
\right)} u \right]}^{1/(2m)}\,,
\ea
which implies that
\ba a=
a(u,v) & = & \sqrt{2u}\exp{\left(
\frac{v}{2} \right)}\,, \\
G = G(u,v) & = & \sqrt{2u}\exp{\left( -\frac{v}{2} \right)},
\ea
when $J=0$. Here, we set
\be
\mu \equiv \frac{6n^2}{(1-3n)^2}\,.
\ee

The substitution of these functions $a = a(u,v)$ and $G = G(u,v)$
into $L$ yields
\be \label{32} L'_J = -6\exp{\left( \frac{m+1}{2n
m}v \right)}\dot{u}\dot{v} - W \exp{(3v)} - 8\pi D
\ee
for $J\neq 0$ (and ${\cal K}=0$), which becomes
\be \label{32bis}
L'_0= - 6 \exp{(2v)}\dot{u}\dot{v} + 3{\cal K}\exp{(v)} -W
\exp{(3v)} - 8\pi D
\ee
for $J=0$ and any ${\cal K}$. By
construction, in both cases $u$ is cyclic for $L'$, and there
exists a non-vanishing constant of motion $\Sigma \equiv -\partial
L'/\partial \dot{u}$ associated to $L'$, which helps to solve the
equations deduced from it. We in fact find
\be \label{34}
\Sigma_J= 6 \exp{\left( \frac{m+1}{2n m}v \right)}
\dot{v}\,,\,\,\,\,\,\Sigma_0 = 6 \exp{(2v)} \dot{v}\,,
\ee
being
$\Sigma_J \rightarrow \Sigma_0$ for $J \rightarrow 0$. Let us also
stress that $L'_J$ and $L'_0$ are non-degenerate, the
related Hessians being non-vanishing.

In order to see the main differences between the case here studied
and the pure gravity one, we have also to consider the
\emph{energy} functions
\ba
E'_J & = & - 6\exp{\left(
\frac{m+1}{2n m}v \right)}\dot{u}\dot{v} + W \exp{(3v)} +
8\pi D\,, \\ E'_0 & = & - 6\exp{(2v)}\dot{u}\dot{v} - 3{\cal
K}\exp{(v)} + W \exp{(3v)} + 8\pi D\,.
\ea
We again find
that $E'_J \rightarrow E'_0$ for $J \rightarrow 0$ and ${\cal K} =
0$, once we note that the presence of ${\cal K}\neq 0$ is indeed
evident only in the expressions of $L$ and $E_L$ and their
transformed forms. (When ${\cal K}\neq 0$, we in fact have to add
the term $- 3{\cal K}\exp{(v)}$ in $E'_0$.)

In the following, we will treat separately the two cases with
$J=0$ and $J\neq 0$. This should make it clearer, on one side, the
contribution of a ${\cal K}$-term and, on the
other side, the range of possibilities tied to choosing different
values of $\mu$ (and, therefore, of $J$ or $n$). This might also
lead to a better understanding of which are the more realistic values that
the parameter $\mu$ should preferably assume from the physical
point of view, also accounting for its nature and origin as a Lagrange
multiplier.

\subsection{$J=0$ and generic ${\cal K}$}

From Eqs. (\ref{34}) and (3.15) (being $E'_0=0$), we find
\be \label{37}
v = v(t) = \frac{1}{2} \ln{\left(\frac{\Sigma_0}{3}t +
2C_1 \right)}\,,
\ee
where $C_1$ is an arbitrary integration constant, and
\be 
\label{38} u = u(t) =
\frac{2\sqrt{3}W (\Sigma_0 t + 6C_1)^{5/2} -
30\sqrt{3}{\cal K}(\Sigma_0 t + 6C_1)^{3/2} + 360\pi D\Sigma_0 t +
45 C_2 {\Sigma_0}^2}{45{\Sigma_0}^2}\,,
\ee
$C_2$ being a further
arbitrary integration constant. From now on, in order to show
simplified formulae, we choose to set $C_2=0$, hence losing the full
generality of our solution. (Such a generality is indeed
guaranteed from the existence of the three integration constants
$C_1$, $C_2$, and $\Sigma_0$, plus the arbitrary constant $D$,
which can be seen as a first integral of the condition $E'_0 =0$.)

Assuming that $\Sigma_0$ is positive, we can rescale time by
defining $\tau \equiv \Sigma_0 t + 6C_1$, so that we find (with
$C_2 = 0$)
\be \label{39}
u(\tau) = \frac{2[{\tau}^{3/2}(W
\tau - 15{\cal K}) + 60 \sqrt{3}\pi D(\tau -
6C_1)]}{15\sqrt{3}{\Sigma_0}^2} \,,
\ee
and
\be \label{39bis}
v(\tau) = \frac{1}{2} \ln{\left(\frac{\tau}{3} \right)}\,.
\ee
From Eqs. (3.7) and (3.8), we thus find
\ba
a=a(\tau) & = &
\frac{2{\tau}^{1/4}\sqrt{\sqrt{3}{\tau}^{3/2}(W \tau -
15{\cal K}) + 180 \pi D(\tau - 6C_1)}}
{3^{5/4}\sqrt{5}\Sigma_0}\,,
\\ G = G(\tau) & = &
\frac{2{\tau}^{-1/4}\sqrt{\sqrt{3}{\tau}^{3/2} (W \tau -
15{\cal K}) + 180 \pi D(\tau - 6C_1)}}{3^{3/4}\sqrt{5}\Sigma_0}\,,
\ea
so that
\be 
\label{3.25}
\Lambda = \Lambda(G(\tau)) =
W G^{-2}(\tau) = \frac{15\sqrt{3}{\Sigma_0}^2 W
{\tau}^{1/2}}{4[\sqrt{3}W {\tau}^{5/2} - 15\sqrt{3}{\cal
K}{\tau}^{3/2} + 180\pi D(\tau - 6C_1)]}\,.
\ee
On the other hand, the Hubble parameter turns out to be
\be \label{42}
H = H(t) \equiv
\frac{\dot{a}(t)}{a(t)} \equiv H(\tau) = 
{\Sigma_0 \over a} \frac{d a}{d\tau}
= \frac{3\Sigma_0[\sqrt{3}{\tau}^{3/2}(W \tau -
10{\cal K}) + 90\pi D(\tau -
2C_1)]}{2\tau[\sqrt{3}{\tau}^{3/2}(W \tau - 105{\cal K}) +
180\pi D(\tau - 6C_1)]}\,.
\ee

In the case of spatially flat universe, Eqs. (3.20), (3.21),
(\ref{3.25}) and (\ref{42}) have to be rewritten with ${\cal K} =
0$, of course. In this case, the expressions so found can also be
deduced from those we are going to write in the next subsection,
once we set $J=0$ therein.

For the time being, let us comment a little on what we have found
above, when $J = 0$ and ${\cal K} \neq 0$. Eq. (3.20) can be more
conveniently written as 
\be a(\tau) =
A\sqrt{{\tau}^{3}-B{\tau}^{2}+C{\tau}^{3/2}}\,,
\ee 
with 
\be A
\equiv \frac{2}{3\Sigma_0}\sqrt{\frac{W}{5}}\,,\,\,\,\,\,
B \equiv \frac{15{\cal K}}{W}\,,\,\,\,\,\, C \equiv
\frac{180 \pi D}{\sqrt{3}W}\,,
\ee 
so that the physically meaningful parameters are 
\be {\cal K} = \frac{3}{4}A^2 B
{\Sigma_0}^2\,,\,\,\,\,\, W = \frac{45}{4} A^2
{\Sigma_0}^2\,,\,\,\,\,\, D = \frac{\sqrt{3}}{16\pi}A^2 C
{\Sigma_0}^2\,.
\ee 
On choosing $\tau_0 \equiv \tau (t_0) = 1$ (being
$t_0$ the present time), we get $a_0 \equiv a(t_0) =
A\sqrt{1-B+C}$ and 
\be 
z(\tau) \equiv \frac{a_0}{a} - 1 =
\frac{\sqrt{1-B+C}}{\sqrt{C{\tau}^{3/2}-B{\tau}^2+{\tau}^3}} -
1\,,\,\,\,\,\, H(\tau) = \frac{\Sigma_0 [3C+2\sqrt{\tau}(3\tau -
2B)]}{4\tau [C+\sqrt{\tau}(\tau - B)]}\,,
\ee 
so that the present
value of the Hubble term is $H_0 = \Sigma_0(6-4B+3C)/(4-4B+4C)$.

On the other hand, we can write Eq. (3.21) as \be G(\tau) =
\frac{A}{\sqrt{3}}\sqrt{{\tau}^2-B\tau+C{\tau}^{1/2}}\,,\ee from
which $G_0 \equiv G(\tau =1) = A\sqrt{1-B+C}/\sqrt{3}$. (Note that
such a value is in general different from the Newtonian one.) With
$D$ from Eq. (3.26), we also have the following expression for the
energy density: 
\be \rho = Da^{-3} =
\frac{\sqrt{3}C{\Sigma_0}^2}{16A\pi
({\tau}^3-B{\tau}^2+C{\tau}^{3/2})^{3/2}}\,.
\ee 
Its present value is therefore $\rho_0 =
\sqrt{3}C{\Sigma_0}^2/[16A(1-B+C)^{3/2}\pi]$. As to the present
matter content of the universe, we can continue to use the above
defined value of $G_0$, and find that 
\be \Omega_0 = \frac{8\pi
G_0 \rho_0}{3{H_0}^2} = \frac{8C(1-B+C)}{3(6-4B+3C)^2} 
\ee 
only depends on $B$ and $C$ (i.e., not on $\Sigma_0$). Of course, it is
possible to adjust such parameters to get a value of $\Omega_0$
not very far from $0.3$. In general, it is in fact possible to see
that there exist suitable values for $B$ and $C$ giving something
not substantially different from what is usually produced in the
$\Lambda$CDM model, for instance. This will be anyway better
outlined in the next section, for the case with ${\cal K}=0$,
where a little more refined comment on such a possibility is
worked out.

\subsection{$J\neq 0$ and ${\cal K}=0$}

On setting again $C_2 = 0$ and $\tau \equiv \Sigma_J t + 6C_1$, we now have
\be \label{44}
v = v(\tau) = \frac{2n}{(6n-1)}\ln{\left[
\frac{6n-1}{12n}\tau \right]}\,,
\ee
and
\be \label{45}
u=u(\tau) = \frac{12^{\frac{6n}{1-6n}}\left[ \frac{6n-1}{n}
\right]^{\frac{12n-1}{6n-1}}nW}{(12n-1){\Sigma_J}^2}
{\tau}^{\frac{12n-1}{6n-1}} + \frac{8\pi D(\tau -
6C_1)}{{\Sigma_J}^2}\,.
\ee
From them, with $C_1 =0$ we get $\tau
= \Sigma_J t$ and $a(0)=0$ (which indeed requires attention; see
Ref. \cite{rub1} for detailed considerations on that), so that we have
\ba 
a = a(\tau) & = & A \left(
B{\tau}^{\frac{6n}{6n-1}} + {\tau}^{\frac{12n}{6n-1}} \right)^n , \\
G = G(\tau) & = & C \left( {\tau}^2 + B {\tau}^{\frac{2-6n}{1-6n}}
\right)^{3n-1}\,,
\ea
where we have defined the constants
\ba
A & \equiv & A(n, W, \Sigma_J) \equiv
12^{\frac{n(1+6n)}{1-6n}}n^{\frac{12n^2}{1-6n}}
(6n-1)^{\frac{12n^2}{6n-1}}(12n-1)^{-n}{W}^n
{\Sigma_J}^{-2n}\,,\\
B & \equiv & B(n, W, D) \equiv {W}^{-1}
\left[ 2^{\frac{3(1-10n)}{1-6n}}(3n)^{\frac{6n}{6n-1}}
(6n-1)^{\frac{1-12n}{6n-1}}(12n-1)\pi D \right]\,,\\
C & \equiv & C(n, W, \Sigma_J) \equiv
(6n-1)^{2(3n-1)}\left[ 12n^2(12n-1) \right]^{1-3n}
{W}^{3n-1} {\Sigma_J}^{2(1-3n)}\,.
\ea

Thus, we can see that such expressions are generalizations of the
ones found in the case with $J = 0$, with the usual caution as to
the terms with ${\cal K}\neq 0$. Also, when $D=0$ we recover the
same results obtained in Ref. \cite{Bona06} for the pure gravity
model. Of course, let us note that above we have had to choose
appropriate values of the $n$ parameter, setting $n \neq -1/6, 0,
1/12, 1/6, 1/3$. This affects the values of the interaction
$\mu$ parameter occurring in the Lagrangian; we must then assume
from the very beginning that our treatment is performed with $\mu
\neq 0, 2/27, 2/3$. (The values $\mu = 0,2/3$ have been already
ruled out, since $J\neq 1, 3/2$.)

Furthermore, let us explicitly exhibit the function $\Lambda =
\Lambda (\tau)$ in the matter--dominated gravity regime of the
universe when $J\neq 0$ ($\Rightarrow n\neq 1/2$ and $\mu \neq 6$)
and ${\cal K}=0$
\be \label{3.34}
\Lambda = \Lambda (\tau; n) =
W G^{\frac{1}{1-3n}} = W C^{\frac{1}{1-3n}} \left(
{\tau}^2+B{\tau}^{\frac{2-6n}{1-6n}} \right)^{-1}\,.
\ee
(When $J=0$ ($\Rightarrow n=1/2$) and ${\cal K}=0$, this reduces to the
$\Lambda$-term found in that case.)

Now the Hubble parameter has the more general expression
\be \label{3.34bis}
H = H(t) \equiv \frac{\dot{a}(t)}{a(t)} \equiv
n\frac{\dot{u}}{u} + \frac{1}{2}\dot{u} \equiv H(\tau) =
\frac{\Sigma_J}{a}\frac{d a}{d\tau} = \frac{H_1 + H_2
{\tau}^{\frac{6n}{6n-1}}}{{\Sigma}_J \tau [H_3 + H_4
{\tau}^{\frac{6n}{6n-1}}]}\,,
\ee
where
\ba
H_1 & \equiv &
H_1 (D) \equiv 48\pi D(12n-1)n^{\frac{2(9n-1)}{6n-1}}\,, \\
H_2 & \equiv & H_2 (W, \Sigma_J) \equiv W n^2
\left[ \frac{(6n-1)^{12n-1}}{12}{\Sigma_J}^{12n}
\right]^{\frac{1}{6n-1}}\,, \\ 
H_3 & \equiv & H_3 (D) \equiv 8\pi
D [1+18n(4n-1)]n^{\frac{6n}{6n-1}}\,, \\ 
H_4 & \equiv & H_4
(W,\Sigma_J) \equiv W \left[
{\Sigma_J}^2(6n-1) \right]^{\frac{6n}{6n-1}} \left(
12^{\frac{6n}{6n-1}}-12^{\frac{-1}{6n-1}}n +
3^{\frac{2(3n-1)}{6n-1}}4^{\frac{-1}{6n-1}}n^2 \right).
\ea

In the next section we will use such formulae to produce a
cosmological model able to be compared successfully with the
concordance $\Lambda$CDM one, which makes it interesting to further study
our model in the future.

\section{The ${\cal K}=0$ model and the observations}

A way to understand whether the cosmological model resulting from
the above expressions for $a$, $G$, and $\Lambda$ may be
considered for further analysis is to study how the results
obtained can fit observational data from SNIa. Usually one should
take at least the Gold Data Set \cite{gold} as observational
result and choose the constants involved above in order to obtain
a best fit. Here, we simply limit ourselves to directly compare
our ${\cal K}=0$ model with the concordance $\Lambda$CDM one, which,
as known, fits those data very well. This should make it possible to
infer a first feasibility of the model here considered. Of course,
further investigations are anyway necessary in order to assess all
related issues and have to be postponed to future work.

\subsection{Some useful cosmological parameters}

First of all, let us limit our attention to the ${\cal K}=0$ case
with $C_2=0$, so that we can use the generic $J$ formulae above.
(Things sensibly change when ${\cal K} \neq 0$ and have to be
analyzed separately.) In order to further simplify the expressions
below, let us also set $\Sigma_J = 1$ and $C_1 =0$ in them, which
arbitrarily fixes the time scale and origin. From $\tau = \Sigma_J
t = t$ we then have $a(0)=0$. Setting $W =1$ requires more
attention, on the other hand, even if the $W$ parameter
only appears as a factor in the expression of $u$. This is, in
fact, a very peculiar situation which will be dismissed in the
last part of the next subsection B, in order to better understand
how well the model defines the time evolution of $G$. Here, we can
proceed rather roughly, because the aim of this subsection is only
that of showing that there exists at least one \emph{reasonable}
choice of the parameters. Thus, in the following let us start by
considering the cosmic scale factor $a = a(t)$ as given by Eq.
(3.24) with \ba A &= &
12^{\frac{n(1+6n)}{1-6n}}n^{\frac{12n^2}{1-6n}}
(6n-1)^{\frac{12n^2}{6n-1}}(12n-1)^{-n}\,,\\
B & = & 2^{\frac{3(1-10n)}{1-6n}}(3n)^{\frac{6n}{6n-1}}
(6n-1)^{\frac{1-12n}{6n-1}}(12n-1)\pi D \,,
\ea
so that fixing the
value of $n\neq -1/6, 0, 1/12, 1/6, 1/3$ simply yields $A$ as a
number, while $B$ still depends on $D$. On the other hand, the
expression for $G$ is given by Eq. (3.27) with
\be \label{new1}
C= (6n-1)^{2(3n-1)}\left[ 12n^2(12n-1) \right]^{1-3n} \,,
\ee
from which, using Eq. (\ref{3.34}), one can get the related expression
of $\Lambda = \Lambda (t)$.

Other relevant quantities for observations are
\ba
H & = & H(t) =
\frac{H_1 + H_2 t^{\frac{6n}{6n-1}}}{ t\left( H_3 + H_4
t^{\frac{6n}{6n-1}} \right)}\,,\\
\Omega_m & = & \Omega_m (t) \equiv \frac{8\pi G D}{3H^2 a^3} =
\frac{32\pi D n u(t)e^{\frac{1-6n}{2n}v(t)}}{3(2n \dot{u}+u
\dot{v})^2}= \frac{{\Omega_m}^{(1)} + {\Omega_m}^{(2)}
t^{\frac{6n}{6n-1}}}{{\Omega_m}^{(3)} + {\Omega_m}^{(4)}
t^{\frac{6n}{6n-1}} + {\Omega_m}^{(5)} t^{\frac{12n}{6n-1}}}\,.
\ea
Here, $H_1$ and $H_3$ are those in Eqs. (3.33) and (3.35), while
$H_2$ and $H_4$ are now
\ba
H_2 & \equiv & H_2 (W =
\Sigma_J = 1) \equiv n^2 \left[ \frac{(6n-1)^{12n-1}}{12}
\right]^{\frac{1}{6n-1}}\,, \\ 
H_4 & \equiv & H_4 (W =
\Sigma_J = 1) \equiv (6n-1)^{\frac{6n}{6n-1}} \left(
12^{\frac{6n}{6n-1}}-12^{\frac{-1}{6n-1}}n +
3^{\frac{2(3n-1)}{6n-1}}4^{\frac{-1}{6n-1}}n^2 \right).
\ea
On the other hand, the constants ${\Omega_m}^{(1)},\,\ldots
\,,{\Omega_m}^{(4)}$ are all depending on $D$, while
${\Omega_m}^{(5)}$ does not depend on it; they are defined as
\ba
{\Omega_m}^{(1)} & \equiv &
2^{\frac{8}{1-6n}}3^{\frac{2(9n-1)}{1-6n}}(1-12n)^2
(6n-1)n^{\frac{2(3n-1)}{1-6n}}{\pi}^2 D\,,
\\ {\Omega_m}^{(2)} & \equiv & 2^{\frac{5(1+6n)}{1-6n}}
3^{\frac{2(12n-1)}{1-6n}} \frac{12n-1}{n^2}
(6n-1)^{\frac{2(9n-1)}{6n-1}}\pi D\,, \\
{\Omega_m}^{(3)} & \equiv &
2^{\frac{8}{1-6n}}3^{\frac{6n}{1-6n}}(1-12n)^2
n^{\frac{6n}{6n-1}}\pi D^2\,, \\{\Omega_m}^{(4)} & \equiv &
2^{\frac{7+18n}{1-6n}}3^{\frac{18n}{1-6n}}
(6n-1)^{\frac{6n}{6n-1}}\left[ 1 + 2\cdot
3^{\frac{2(1-9n)}{1-6n}}n(4n-1) \right]\pi D\,,\\ {\Omega_m}^{(5)}
& \equiv &
2^{\frac{4(12n+1)}{1-6n}}3^{\frac{18n}{1-6n}}n^{\frac{6n}{1-6n}}
(6n-1)^{\frac{2(12n-1)}{6n-1}}\,.
\ea
We can also introduce the
parameters
\ba
\Omega_{\Lambda} & = & \Omega_{\Lambda} (t) \equiv
\frac{\Lambda}{3H^2}= \frac{4n u(t)e^{\frac{v(t)}{2n}}}{3(2n
\dot{u}+u \dot{v})^2} =
\frac{{\Omega_{\Lambda}}^{(1)}t^{\frac{6n}{6n-1}} +
{\Omega_{\Lambda}}^{(2)}
t^{\frac{12n}{6n-1}}}{{\Omega_{\Lambda}}^{(3)} +
{\Omega_{\Lambda}}^{(4)} t^{\frac{6n}{6n-1}} +
{\Omega_{\Lambda}}^{(5)} t^{\frac{12n}{6n-1}}}\,,\\\Omega_G & = &
\Omega_G (t) \equiv \frac{\mu {\dot{G}}^2}{6H^2 G^2}= \frac{(- 2n
\dot{u}+u \dot{v})^2}{(2n \dot{u}+u \dot{v})^2} =
\frac{({\Omega_G}^{(1)} + {\Omega_G}^{(2)}
t^{\frac{6n}{6n-1}})^2}{{\Omega_G}^{(3)} + {\Omega_G}^{(4)}
t^{\frac{6n}{6n-1}} + {\Omega_G}^{(5)} t^{\frac{12n}{6n-1}}}\,,
\ea
where
\ba
{\Omega_{\Lambda}}^{(1)} & \equiv & \left[
2^{18n-1}(3n)^{6n}(6n-1) \right]^{\frac{1}{6n-1}}[1+18(4n-1)]^2
\pi D\,, \\
{\Omega_{\Lambda}}^{(2)} & \equiv & \frac{1}{4}
(2n-1)(6n-1)^{\frac{2(12n-1)}{6n-1}}\,, \\
{\Omega_{\Lambda}}^{(3)} & \equiv & \left[
2^{\frac{2(12n-1)}{6n-1}}(3n)^{\frac{12n-1}{6n-1}}
(1+12n) \right]^2 {\pi}^2 D^2\,, \\
{\Omega_{\Lambda}}^{(4)} & \equiv & \left[ 2^{3(10n-1)}
(3n)^{2(9n-1)}(6n-1)^{12n-1}\right]^{\frac{1}{6n-1}}(12n-1)\pi D\,,\\
{\Omega_{\Lambda}}^{(5)} & \equiv & \left[
3n(6n-1)^{\frac{12n-1}{6n-1}}\right]^2\,,
\ea
(only ${\Omega_{\Lambda}}^{(2)}$ and ${\Omega_{\Lambda}}^{(5)}$
being independent of $D$) and
\ba
{\Omega_G}^{(1)} & \equiv & -
\left(\frac{2^4 3^{12n}}{n^{6n}}
\right)^{\frac{1}{1-6n}}[1+3n(12n-5)]\pi D\,,
\\ {\Omega_G}^{(2)} & \equiv & \left[ \frac{6^{18n}}{2(6n-1)}
\right]^{\frac{1}{1-6n}}[1-18n(1+6n(2n-1))]\,, \\
{\Omega_G}^{(3)} & \equiv & \left[
2^{3(7-12n)}3^{2(1-3n)}n^{2(1-12n)}
\right]^{\frac{1}{1-6n}}(1-12n)^2 {\pi}^2 D^2\,,
\\{\Omega_G}^{(4)} & \equiv & \left[ \frac{2^{7-18n}3^2
n^{2(1-9n)}}{(6n-1)^{6n}}\right]^{\frac{1}{1-6n}}
[1+18n(4n-1)]\pi D\,,\\
{\Omega_G}^{(5)} & \equiv & \left[ 12^{1+3n}(6n-1)^{1-12n}
\right]^{\frac{2}{1-6n}}n^2 ,
\ea
with, again, only
${\Omega_G}^{(2)}$ and ${\Omega_G}^{(5)}$ independent of $D$.

\subsection{Comparison with the $\Lambda$CDM model}

In order to begin to compare the model characterized by the
parameters above with the usual $\Lambda$CDM model, let us
consider the behaviour of the scale factor $a$ for large and small
times. In the former case, this is equivalent to take the
expression in Eq. (3.26) with $B \rightarrow 0$, $a_{t \rightarrow
\infty} \equiv At^{\frac{12n^2}{6n-1}}$, while in the latter 
we can assume $a_{t \rightarrow 0} \equiv AB^n
t^{\frac{6n^2}{6n-1}}$. This means that the two exponents
characterizing the asymptotic time behaviours of the scale factor
are respectively
\be
p_1 \equiv \frac{12n^2}{6n-1}\,,\,\,\,\,\,
p_2 \equiv \frac{6n^2}{6n-1},
\ee
where we want that $p_1 >
1$ and $p_2 < 1$, which implies that we find a limited range of
variability for the $n$ parameter, $(3-\sqrt{3})/6 < n <
(3+\sqrt{3})/6$. This gives rise to a constraint on the values of
the $\mu$ parameter introduced in the Lagrangian at the beginning
of our investigation, since $\mu$ is indeed equal to $2$ at both
extremals, but we also find that $\mu \rightarrow \infty$ when $n
\rightarrow 1/3$. Thus, the range of values of $\mu$ is indeed
open from above, i.e. $\mu > 2$.

Let us now consider the behaviour of $G=G(t)$ for large and small
times. As before, this yields $G_{t \rightarrow \infty} \equiv
Ct^{2(3n-1)}$ (for $B \rightarrow 0$) and $G_{t
\rightarrow 0} \equiv CB^{3n-1} t^{\frac{2(1-3n)^2}{6n-1}}$. 
The two exponents involved are now
\be
p_3 \equiv
2(3n-1)\,,\,\,\,\,\, p_4 \equiv \frac{2(1-3n)^2}{6n-1}\,.
\ee
Here, if we want that both exponents do not attain too high a value,
we can stick around $n = 1/3$, which, as already above, represents
a degenerate situation and cannot be considered exactly in our
analysis. Anyway, let us underline that, for $n \rightarrow 1/3$,
we get
\be
a(t) \rightarrow A\left[ t^2(B + t^2)
\right]^{\frac{1}{3}}\,,\,\,\, H(t) \rightarrow \frac{2B +
4t^2}{3Bt + 3t^3}\,,\,\,\, G(t) \rightarrow C\,,
\ee
which recovers (once we set $B=1$) the solution found in Ref. \cite{rub1},
already known to fit observational SnIa data \cite{rub1,rub2}. In
general, however, $H = H(t)$ is given by Eq. (4.4) and can be
written as
\be
H(t) = \frac{6n^2 \left( B + 2t^{\frac{6n}{6n-1}}
\right)}{(6n-1)t\left( B + t^{\frac{6n}{6n-1}} \right)}\,.
\ee
Once we choose the present time $t_0 =1$, the present scale factor and
Hubble term respectively become
\be
a_0 = A(1 + B)^n\,,\,\,\,\,\,
H_0 = \frac{6(2+B)n^2}{(1+B)(6n-1)}\,,
\ee
so that in such units
$H_0 \sim 1$. In what follows, we therefore choose to set $H_0 =
1$ exactly, even if this has to be considered as simply heuristic,
since we have already fixed the time scale by setting $\Sigma_J =
1$ and the age of the universe with $t_0 = 1$. This makes things
analytically simpler but, at the same time, underlines the fact
that we have chosen to consider the model only at a first
glance, without even trying to fit observational data, just to
look at a first viability of our model. A more correct analysis
then deserves more and better work, of course. Here, we have
chosen to focus our considerations on the analytical method
adopted to get the model itself rather than on its compatibility with
observations.

If $H_0 = 1$ one finds that
\be
B =\frac{-1+6n-12n^2}{1-6n+6n^2}\,,
\ee
which inserted into the expression of the Hubble term leads to
\be
H = \frac{6n^2 \left(
\frac{-1+6n-12n^2}{1-6n+6n^2} + 2t^{\frac{6n}{6n-1}} \right)
}{(6n-1)t\left( \frac{-1+6n-12n^2}{1-6n+6n^2} +
t^{\frac{6n}{6n-1}} \right)}\,.
\ee
Let us then note that we have
not chosen to normalize $a_0$, and that, using such a value for
$B$, the definition of the redshift $z \equiv a_0/a - 1$ leads to
\be
z = 1 +  \left( -\frac{6n^2}{1-6n+6n^2}\right)^n \left[
\frac{(-1+6n-12n^2)t^{\frac{6n}{6n-1}}}{1-6n+6n^2} +
t^{\frac{12n}{6n-1}} \right]^{-n}\,.
\ee
In this way, we see that the only parameter to be considered is $n$.

Let us now compare the function $H = H(z)$ we can deduce from such
expressions (where we set, for example, $n = 0.3$) with that
usually estimated in a $\Lambda$CDM model with $\tilde{\Omega}_m
=0.27$ 
\be \tilde{H}_{\Lambda}=\sqrt{1 - \tilde{\Omega}_m +
\tilde{\Omega}_m(1 + z^2)^3}\,, 
\ee 
hence finding a first but
apparently not very good agreement, as shown in Fig. 1.
\begin{figure} 
\centering \resizebox{8.5cm}{!}
{\includegraphics{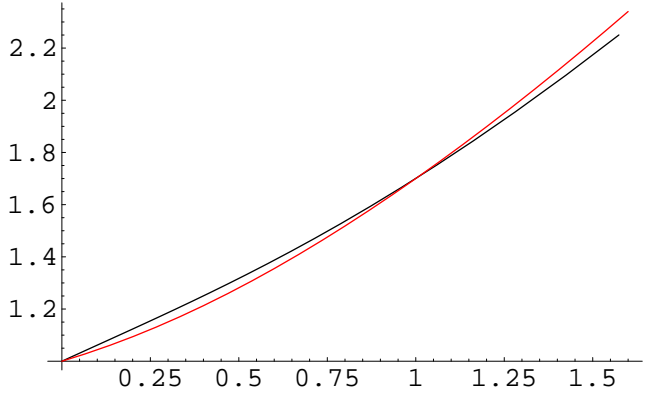}} \caption{Comparison between the Hubble
terms $H=H(z)$ in our model (red) and $\tilde{H}=\tilde{H}(z)$ in
the $\Lambda$CDM model (black).} \label{exponent}
\end{figure}
(Notice that the tilde parameters are those peculiar to the
$\Lambda$CDM model, which can of course be different from the ones
in our own model.) As a matter of fact, however, using the
Hubble--free luminosity distance $d_L$ gives a plot (see Fig. 2)
with a very good agreement, at least for $z < 2$.
\begin{figure} 
\centering \resizebox{8.5cm}{!}
{\includegraphics{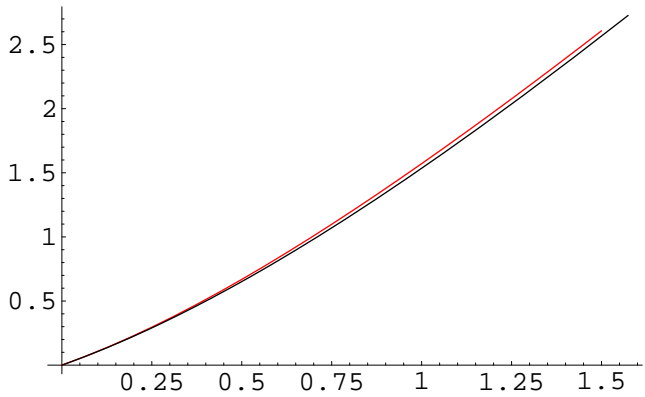}} \caption{Comparison betwwen the
luminosity distances $d_L=d_L(z)$ in our model (red) and
$\tilde{d_L}=\tilde{d_L}(z)$ in the $\Lambda$CDM model (black).}
\label{exponent}
\end{figure}

To understand better what goes on, let us first of all stress that
$\Sigma_J = 1$ and $t_0 = 1$ tell us that we are using the age of
the universe as unit. On the other hand, if we now dismiss our
previous choice for $W$, write the expression of $B$
without setting $W = 1$, and compare it with the one we
have found as a result of $H_0 =1$, we get $1.077 = 2409.08
D/W$, from which $D = 0.00045 W$. Making something
similar with the expression of $C$ analogously leads to $C =
1.16/W^{0.1}$. In this way, it turns out that the present
value of $G = G(t)$ is
\be
G_0 \equiv G(t_0 = 1) =
\frac{1.08}{{W}^{0.1}}\,.
\ee
Of course, we do not
expect that $G_0 \equiv G_N \equiv G_{\rm Newton}$, even if we can
determine $W$ in order to get it. We can take $W =
1$ or, for example, $G_0 = G_N = 1$; in this latter case, we find
$W = 2.11$. Anyway, we have to consider that small
differences between $G_0$ and $G_N$ make this value vary a lot. To
better investigate this issue, we can evaluate the fractional time
rate of change of $G$
\be
\frac{\dot{G}}{G_0}= -8.33 \times 10^{-2}\,.
\ee
Here, it is important to remember that we are using the age of the
universe as unit, so that the effective rate is of the order
$10^{-13}$ yr$^{-1}$. We can also estimate how $G$ (with $n =
0.3$) varies from now ($z_0 = 0$) up to the equivalence time
($z_{eq} \sim 10000$), by exploiting the formula
\be
G(t) =\frac{1.076}{\frac{1.077}{t^{0.25}}+t^2}\,.
\ee
This shows that the variation is $30\%$ and can be seen as acceptable
in such first rough considerations. On the other hand, the infrared
fixed-point hypothesis is that the nontrivial running is due to quantum
fluctuations with momenta smaller than $k_{\rm cosmo}$, where
$1/k_{\rm cosmo}$ is the length scale characterizing the largest localized 
structures in the Universe. There is no conflict with the classical tests
of general relativity \cite{Will05}.

Looking at the present value of $\Lambda$, we find $\Lambda_0
\equiv \Lambda (t_0 =1) = 2.11$, which is indeed in ${H_0}^2$
units. As to the more fundamental test on $\Omega_m$, we have
\be
\Omega_{m0} \equiv \frac{8\pi G_0 D}{3{a_0}^3} = \frac{2.09 \times 10^{-3}
{W}^{0.9}}{A^3}\,.
\ee
Since $A = 2.07 \times 10^{-1} \times 
{W}^{0.93}$, we get
\be
\Omega_{m0} = 2.33 \times 10^{-1}\,,
\ee
which is indeed a very reasonable value, because all our
comparison procedure was meant to emulate what done with the
solution obtained in Refs. \cite{rub1,rub2}. (Notice that this
result does not depend on $W$.)

We can also evaluate
\be
\Omega_{\Lambda 0} \equiv
\frac{\Lambda_0}{3} = 7.04 \times 10^{-1}\,,\,\,\,\,\, \Omega_{G 0} \equiv
\frac{\mu {\dot{G}}^2}{6G^2}(t = t_0) = 6.25 \times 10^{-2}\,,
\ee
from which we deduce the usually expected constraint
\be
\Omega_{m0}+\Omega_{\Lambda 0}+\Omega_{G 0} = 1\,.
\ee

\section{Conclusions}

Since a renormalization-group approach seems to show that quantum Einstein
gravity \cite{Reut98} is asymptotically safe \cite{Laus02b} 
despite being non-renormalizable at 
perturbative level, we began investigating the cosmological applications
in Ref. \cite{Bona04}, devoted to the Lagrangian and Hamiltonian
formulation with variable $G$ and $\Lambda$. In the present paper, we have 
solved the equations from an RG-improved gravity Lagrangian for a homogeneous
and isotropic, matter--dominated universe, finding the related
dynamical behaviour for the gravitational coupling $G$ and the
cosmological $\Lambda$-term. We have used the so-called
\emph{Noether Symmetry Approach}, yielding a coordinate
transformation which leads to a form of the equations easily and
exactly solvable \cite{deritis90,cap96,Bona06}. This has required
to impose the existence of a Noether symmetry for the point--like
Lagrangian describing the cosmological dynamics. The existence of
the coordinate transformation so found, which might indeed also be
guessed a priori, is a direct product of the machinery worked out
here even if the solutions we obtain are independent of the very
existence of the symmetry. On the other hand, as we have already
found in the pure gravity case \cite{Bona06}, such a symmetry
brings into our treatment some strong conditions, like those on
the values of ${\cal K}$ and the $J$ parameter, and on the form of
the function $\Lambda = \Lambda (G)$.

While in Ref. \cite{Bona04} a power-law behaviour for the scale factor
was guessed from the beginning to solve the equations, we have
seen already in Ref. \cite{Bona06} that this can instead
result exactly and generally from the method itself. When ${\cal
K}=0$, in Ref. \cite{Bona04} it was assumed $a=At^{\alpha}$ for the
scale factor; with arbitrary $A$, this gives $\alpha_{\pm} = (3\pm
\sqrt{9+12{\xi}^2 \lambda_{\star}})/6$, in close connection with
the hypothesis of being in the neighbourhood of a fixed point
$(g_{\star},\lambda_{\star})$, which constrains $G$ and $\Lambda$
to be
\be \label{3.35}
G = G(t) =
g_{\star}{\xi}^{-2}t^2\,,\,\,\,\,\, \Lambda = \Lambda (t) =
\lambda_{\star}{\xi}^2 t^{-2}\,\,\,\, \Rightarrow \,\,\,\,
G\Lambda = g_{\star}\lambda_{\star} \equiv {\rm constant} \,.
\ee
This allows arbitrarily large values for ${\alpha}_{+}$ (since $\xi$ is
undetermined), and constrains the interaction parameter to be
$\mu_{\pm} = 3\alpha_{\pm}/2 = (3\pm \sqrt{9+12{\xi}^2
\lambda_{\star}})/4$ \cite{Bona04}. If $\lambda_{\star} > 0$,
this leads to power-law inflation for the ``+" solution.

Under such assumptions it is possible, as in Ref. \cite{Bona06}, to
see that the general asymptotic trend of $a$ for large $\tau$ is
power-law, $a \sim {\tau}^p$ (with $p \equiv (3-2J)^2/[3(J^2 -3J
+2)]$), without strictly imposing to be near a fixed point, but
rather being probably very far from it. We again find acceleration
only when $J<1$ or $J>2$, $p$ being always $>1$ in these ranges;
when $J$ is near the values $1$ and $2$, the exponent can assume
any large value, therefore yielding a possible strong power-law
behaviour. On the other hand, the interaction parameter $\mu$ is a
function of $J$ such that both $p$ and $\mu$ assume symmetric
values for $J<1$ and $J>2$. We also find a power--law behaviour,
with exponent $p^{\prime} \equiv 1/(1-J)$, for the function $G =
G(\tau)$. A special case is obtained when $p^{\prime} =2$,
that is $J=1/2$; here, we have an accelerated stage, with
$p=16/9$. In the $J=0$ case, $p$ is instead fixed to be $3/2$,
with a soft acceleration for the universe.

It is interesting to note that the model with ${\cal K} \neq
0$ also makes it possible to perform a sort of rough comparison
with the $\Lambda$CDM model, even if more refined work has to be
done in such a direction. Here, in fact, we have limited ourselves
to outline the procedure adopted to find solutions rather than the
necessary comparison with observation, which is then postponed to
future work.

In general, we have to stress that the presence of an ordinary
matter (dust) term in the theory makes larger differences with
respect to what is found in the proximity of the non-Gaussian
ultraviolet fixed point, and also with respect to the pure gravity
case (which can be probably interpreted as characterizing the
epoch soon outside the attraction basin of that fixed point), and
can be a little better understood by means of the solutions
resulting from the cosmological equations. As a matter of fact, we
get power-law dependence on time for the scale factor, which
might also allow for accelerated trends. Of course, these would be now
set up in a period next to the much earlier one characterized by a
pure gravity regime, in which standard inflation was born
\cite{Bona06}. This could mean that what we have effectively
treated here regards a cosmic evolutionary stage close to the
current one, this latter being first assumed (and tested) as
matter--dominated and then recently discovered as an accelerated
one. In our context, quantum effects induce running of both the
gravitational coupling $G = G(t)$ and the cosmological term
$\Lambda = \Lambda (t)$. This in fact yields primordial inflation
soon after the universe exits the region where the attraction basin
of the non-Gaussian (ultraviolet) fixed point works
\cite{Bona06}, as well as another inflationary behaviour in or,
probably, after a matter--dominated epoch, and always without need
to introduce a scalar field in the cosmic content. The usual
matter domination should indeed occur in between, also after what
is usually described as radiation era. This means that we do not
have any concrete continuous model of the evolution of the
universe, while also assuming that our considerations might be
valid once we accept an onset of quantum effects only at the
beginning and at the end of such evolution.

What also lacks in this sort of a patchwork reconstruction of
cosmological history mainly concerns the way one could establish
suitable links among these separated stages. Nonetheless, if we do
accept the very recently assumed existence of another
non-Gaussian, infrared fixed point late in the history of the
universe, it could indeed characterize a later inflationary
period. (This last regime of the evolution of the universe,
typically pictured as an asymptotic dark-energy dominated
stage, is mostly described by means of a scalar field. Let us
again stress, here, that in our treatment we do not need to
introduce such an ingredient in the cosmic content.) Thus, other
important physics would be needed, linking this later era to the
previous ones, which prevents a thorough understanding of the
whole picture we are drawing. As a result, however, we would be
partially facing the construction of a global patchwork model for
the cosmic evolution which seems to be of some interest, because,
once we introduce the quantum effects in a RG--inspired effective
theory of gravity, we do not need any other mechanism to work out
accelerated stages.

Still lacking is also the concrete possibility to generate an
acceptable radiation--dominated regime, which is instead needed to
describe what we understand as the usual standard cosmological
model, in the framework of RG--inspired cosmology. The procedure
adopted in this paper, using the Noether Symmetry Approach, does
not in fact easily work with a Lagrangian where the matter term is
$L_m \equiv Da^{-1}$ (since $\gamma = 4/3$ for radiation). Thus,
other methods have to be worked out for treating the problem of
solving cosmological equations in such a period. Finally, it
should be stressed that, in the context of our paper, the
development of a perturbation theory able to produce suitable
structures in the universe is meaningless, since it can indeed
work only in an \emph{appropriate} matter--dominated era, when the
scale factor is $a \sim t^{2/3}$, which we never find, instead, in
our treatment.

\acknowledgments
The authors are grateful to the INFN for
financial support. The work of G. Esposito has been partially
supported by PRIN {\it SINTESI}.

\end{document}